\documentclass[12pt]{iopart}
\usepackage[dvips]{graphicx}

\begin{document}

\title{Energy Spectrum for Neutral Collective Excitations 
in Striped Hall States}
\author{N Maeda, T Aoyama, Y Ishizuka, and K Ishikawa}
\address{Department of Physics, Hokkaido University, 
Sapporo 060-0810, Japan}
\begin{abstract}
Neutral collective excitations in the striped Hall state are studied by 
using the single mode approximation and Hartree-Fock approximation 
at the half-filled third and fourth Landau level. 
We find that the spectrum includes anisotropic NG modes and 
periodic line nodes. 
\end{abstract}

\section{Introduction}
In recent experiments, highly anisotropic states, which have 
anisotropic longitudinal resistances, were observed around the 
half-filled third and higher Landau levels\cite{Lilly,Du}. 
It is believed that the anisotropic state is the striped Hall 
state which is a unidirectional charge density wave in the mean field 
theory\cite{Kou,Moe}. 
The anisotropy of the resistance is naturally explained by the 
anisotropic Fermi surface in the magnetic Brillouin zone\cite{Imo,Mae,Ao}. 

In the absence of disorder and edges, 
the quantum Hall system has the magnetic translation and rotation symmetry. 
In the striped state, a magnetic translation in one direction is 
spontaneously broken to the discrete translation and the rotation is also 
spontaneously broken to the $\pi$-rotation. 
Using the conserved current, the property of the neutral collective 
excitations is studied. 
The spectrum of the neutral collective excitation is obtained in 
the single mode approximation numerically\cite{SMA}. 
The single mode approximation is useful for the fractional 
quantum Hall state (FQHS). 
We show that the spectrum fot the striped Hall state 
has a multiple line node structure and cusps. 
Furthermore, the spectrum has anisotropic feature, that is, 
in one direction the spectrum resembles the liquid Helium 
spectrum with the phonon and roton minimum, and in another 
direction it resembles the FQHS spectrum. 

\section{Symmetries}

Let us consider the two-dimensional electron system in a uniform magnetic 
field $B=\partial_x A_y-\partial_y A_x$. 
We ignore the spin degree of freedom and use the natural unit 
($\hbar=c=1$) in this paper. 
We introduce two sets of coordinates, relative coordinates and 
guiding center coordinates. 
The relative coordinates are defined by 
\begin{equation}
\xi={1\over eB}(-i\partial_y+eA_y),\ 
\eta=-{1\over eB}(-i\partial_x+eA_x).
\end{equation}
The guiding center coordinates are defined by 
\begin{equation}
X=x-\xi,\ Y=y-\eta.
\end{equation}
These coordinates satisfy the following commutation 
relations, 
\begin{eqnarray}
&[X,Y]=-[\xi,\eta]=i/eB,\nonumber\\
&[X,\xi]=[X,\eta]=[Y,\xi]=[Y,\eta]=0.
\end{eqnarray}
The operators $X$ and $Y$ are the generators of the magnetic 
translations of the one-particle state in $-y$ direction and 
$x$ direction respectively. 
The angular momentum $J$ is written as 
\begin{equation}
J={eB\over2}(\xi^2+\eta^2-X^2-Y^2).
\end{equation}
$J$ is the generator of the rotation of the one-particle state. 

The total Hamiltonian $H$ for the interacting charged particles 
is the sum of the free Hamiltonian $H_0$ and the Coulomb interaction 
Hamiltonian $H_{\rm int}$ as follows,
\begin{eqnarray}
H&=&H_0+H_{\rm int},\nonumber\\
H_0&=&\int d{\bf r} \Psi^\dagger({\bf r}){m \omega_c^2\over2}
(\xi^2+\eta^2)\Psi^\dagger({\bf r}),\\
H_{\rm int}&=&{1\over2}\int d{\bf r}d{\bf r}'\Psi^\dagger({\bf r})
\Psi^\dagger({\bf r}')V({\bf r}-{\bf r}')\Psi({\bf r}')\Psi({\bf r}),
\nonumber
\end{eqnarray}
where $\Psi$ is the electron field operator, $\omega_c=eB/m$ and 
$V({\bf r})=q^2/r$ 
($q^2=e^2/4 \pi \epsilon $, $\epsilon$ is the 
dielectric constant) for the Coulomb potential. 

Conserved charges are obtained as the spatial integral of the zeroth 
component of the Noether currents for the symmetries. 
We define the conserved charges, $Q$, $Q_X$, $Q_Y$, and $Q_J$ for 
U(1), magnetic translations, and rotation symmetry respectively, 
as follows, 
\begin{eqnarray}
&Q=\int j^0({\bf r})d{\bf r},&\nonumber\\
&Q_X=\int j_X^0({\bf r})d{\bf r},\ 
Q_Y=\int j_Y^0({\bf r})d{\bf r},&\\
&Q_J=\int j_J^0({\bf r})d{\bf r}.&\nonumber
\end{eqnarray}
Noether currents are defined by 
\begin{eqnarray}
j^\mu&=&{\rm Re}(\Psi^\dagger v^\mu\Psi),\nonumber\\
j_X^\mu&=&{\rm Re}(\Psi^\dagger v^\mu X \Psi)-{1\over eB}\delta_y^\mu
{\cal L},\\
j_Y^\mu&=&{\rm Re}(\Psi^\dagger v^\mu Y \Psi)+{1\over eB}\delta_x^\mu 
{\cal L},\nonumber\\
j_J^\mu&=&{\rm Re}(\Psi^\dagger v^\mu J \Psi)+
\epsilon_{0\mu i}x^i{\cal L},\nonumber
\end{eqnarray}
where $v^\mu=(1,{\bf v})$, ${\bf v}=\omega_c(-\eta,\xi)$, 
Re means real part, and $\cal L$ is the 
Lagrangian density for the total Hamiltonian $H$. 
These charges commute with the total Hamiltonian $H$ and 
obey the following commutation relations, 
\begin{eqnarray}
&[Q_X,Q_Y]&={i\over eB}Q,\nonumber\\ 
&[Q_J,Q_X]&=iQ_Y,\\ 
&[Q_J,Q_Y]&=-iQ_X.\nonumber
\end{eqnarray}
The U(1) charge $Q$ commutes with all conserved charges. 
We assume that $Q$ is not broken and the ground state is the eigenstate 
of $Q$ as $Q\vert 0\rangle=N_e\vert 0\rangle$ where $N_e$ is a number of 
electrons. 

The commutation relations between the conserved charges and 
the current density operators read,
\begin{eqnarray}
&[Q_X,j^\mu]&=-{i\over eB}\partial_y j^\mu,\ 
[Q_Y,j^\mu]={i\over eB}\partial_x j^\mu,\nonumber\\ 
&[Q_X,j_X^\mu]&=-{i\over eB}\partial_y j_X^\mu,\ 
[Q_Y,j_X^\mu]={i\over eB}\partial_x j_X^\mu-{i\over eB}j^\mu,
\nonumber\\ 
&[Q_X,j_Y^\mu]&=-{i\over eB}\partial_y j_Y^\mu+{i\over eB}j^\mu,\ 
[Q_Y,j_Y^\mu]={i\over eB}\partial_x j_Y^\mu,\label{com}\\
&[Q_X,j_J^\mu]&=-{i\over eB}\partial_y j_J^\mu-
ij_Y^\mu,\nonumber\\
&[Q_Y,j_J^\mu]&={i\over eB}\partial_x j_J^\mu+ij_X^\mu\nonumber\\
&[Q_J,j^\mu]&=i(x\partial_y-y\partial_x)j^\mu+i\epsilon_{0\mu\nu}
j^\nu.\nonumber
\end{eqnarray}
The U(1) charge $Q$ commutes with all Noether currents. 
If the expectation value of the right hand side of these equation 
is not zero, corresponding symmetry is spontaneously broken. 

We use the von Neumann lattice (vNL) base for the one-particle 
states\cite{Ima}. 
A discrete set of coherent states of guiding center coordinates, 
\begin{eqnarray}
&(X+iY)\vert\alpha_{mn}\rangle=z_{mn}\vert\alpha_{mn}\rangle,\\
&z_{mn}=a(mr_s+i{n\over r_s}),\ m,\ n;{\rm\ integers},\nonumber
\end{eqnarray}
is a complete set of the $(X,Y)$ space. 
These coherent states are localized at the position $a(mr_s,n/r_s)$, 
where a positive real number $r_s$ is the asymmetric parameter 
of the unit cell and $a=\sqrt{2\pi/eB}$. 
By Fourier transforming these states, we obtain the orthonormal 
basis in the momentum representation
\begin{eqnarray}
&\vert\beta_{\bf p}\rangle=\sum_{mn} e^{ip_x m+ip_y n}
\vert\alpha_{mn}\rangle/\beta({\bf p}),\\
&\beta({\bf p})=(2{\rm Im}\tau)^{1/4}e^{{i\tau p_y^2\over4\pi}}
\vartheta_1({p_x+\tau p_y\over 2\pi}\vert\tau),\nonumber
\end{eqnarray}
where $\vartheta_1$ is a Jacobi's theta function and $\tau=ir_s^2$. 
The two-dimensional lattice momentum $\bf p$ is defined in the 
Brillouin zone (BZ) $\vert p_i\vert<\pi$. 
Next we introduce a complete set in $(\xi,\eta)$ space, that is 
the eigenstate of the one-particle free Hamiltonian, 
\begin{equation}
{m\omega^2_c\over2}(\xi^2+\eta^2)\vert f_l\rangle=
\omega_c(l+{1\over2})\vert f_l\rangle. 
\end{equation}
The Hilbert space is spanned by the direct product of these eigenstates 
\begin{equation}
\vert l,{\bf p}\rangle=\vert f_l\rangle\otimes\vert\beta_{\bf p}\rangle,
\end{equation}
where $l$ is the Landau level index and $l=0$, 1, 2, $\cdots$. 
We set $a=1$ $(eB=2\pi)$ in the following calculation for simplicity. 

Electron field operator is expanded by the vNL base as
\begin{equation}
\Psi({\bf r})=\sum_{l=0}^{\infty}\int_{\rm BZ}{d^2p\over(2\pi)^2}
b_l({\bf p})\langle {\bf r}\vert l,{\bf p}\rangle,
\end{equation}
where $b_l$ is the anti-commuting annihilation operator. 
$b_l({\bf p})$ obeys the non-trivial boundary condition, 
$b_l({\bf p}+2\pi{\bf N})=e^{i\phi(p,N)}b_l({\bf p})$, 
where $\phi(p,N)=\pi(N_x+N_y)-N_y p_x$ and 
${\bf N}=(N_x,N_y)$ are intergers. 
The Fourier transform of the density $\rho({\bf k})=
\int d{\bf r}e^{i{\bf k}\cdot{\bf r}}j^0({\bf r})$ is written as
\begin{equation}
\rho({\bf k})=\sum_{ll'}
\int_{\rm BZ}{d^2p\over(2\pi)^2}b_l^\dagger({\bf p})
b_{l'}({\bf p}-\hat{\bf k})\langle f_l\vert e^{ik_x\xi+ik_y\eta}\vert 
f_{l'}\rangle 
e^{-{i\over4\pi}{\hat k}_x(2p_y-{\hat k}_y)},
\end{equation}
where $\hat{\bf k}=(r_s k_x,k_y/r_s)$. 
Conserved charges are written in the momentum representation as
\begin{eqnarray}
Q&=&\sum_l\int_{\rm BZ}{d^2p\over(2\pi)^2}b_l^\dagger({\bf p})
b_l({\bf p}),\nonumber\\
Q_X&=&r_s\sum_l\int_{\rm BZ}{d^2p\over(2\pi)^2}b_l^\dagger({\bf p})
(i{\partial\over\partial p_x}-{p_y\over2\pi})
b_l({\bf p}),\nonumber\\
Q_Y&=&{1\over r_s}\sum_l\int_{\rm BZ}{d^2p\over(2\pi)^2}b_l^\dagger({\bf p})
i{\partial\over\partial p_y}b_l({\bf p}),
\label{mom}\\
Q_J&=&\sum_l\int_{\rm BZ}{d^2p\over(2\pi)^2}b_l^\dagger({\bf p})
[l+{1\over2}
+\pi\{r_s^2 (i{\partial\over\partial p_x}-{p_y\over2\pi})^2+
r_s^{-2}(i{\partial\over\partial p_y})^2\}]b_l({\bf p}).\nonumber
\end{eqnarray}
As seen in Eq.~(\ref{mom}), 
the magnetic translation in the real space is equivalent to 
the magnetic translation in the momentum space. 
The free kinetic energy is quenched in a magnetic field and 
the one-particle spectrum becomes flat. 
Therefore the free system in a magnetic field is 
translationally invariant in the momentum space. 
We show that there exists the Fermi surface in the mean field 
solution for the striped Hall state. 
The existence of the Fermi surface indicates the violation 
of the translational symmetry in the momentum space or 
in the real space. 

\section{Mean field solution}

The mean field state is constructed as
\begin{equation}
\vert0\rangle=N_1 \prod_{{\bf p}\in {\rm F.S.}}
b_l^\dagger({\bf p})\vert{\rm vac}\rangle,
\label{mfs}
\end{equation}
where F.S. means Fermi sea, $N_1$ is a normalization constant, and 
$\vert{\rm vac}\rangle$ is the vacuum state in which the $l-1$ th and 
lower Landau levels are fully occupied. 
The mean field of the two-point function for the electron field 
is given by
\begin{equation}
\langle0\vert b^\dagger_l({\bf p})b_{l'}({\bf p}')
\vert0\rangle=\delta_{ll'}\theta(\mu-\epsilon_{\rm HF}({\bf p}))
\sum_N(2\pi)^2\delta({\bf p}-{\bf p}'+2\pi {\bf N})e^{i\phi(p,N)},
\end{equation}
where $\mu$ is a chemical potential and one-particle energy 
$\epsilon_{\rm HF}({\bf p})$ is determined self-consistently in the 
Hartree-Fock approximation\cite{Imo,Mae}.  
We assume that the magnetic field $B$ is so strong that 
Landau level mixing effects can be neglected. 
Then we use the Hamiltonian projected to the $l$ th Landau level 
\begin{eqnarray}
P_l H P_l&=&\omega_c N_e^* (l+{1\over2})+ H^{(l)},\\
H^{(l)}&=&P_l H_{\rm int}P_l,\nonumber
\nonumber
\end{eqnarray}
where $P_l$ is the projection operator to the $l$ th Landau level and 
$N_e^*$ is a number of electrons occupying the $l$ th Landau level. 
The free kinetic term is quenched and the total Hamiltonian is reduced 
to the Coulomb interaction term projected to the $l$ th Landau level, 
$H^{(l)}$. 

It was shown that the mean field state with Fermi sea $\vert p_y\vert<\pi 
\nu_*$ satisfies the self-consistency equation at the filling factor 
$\nu=l+\nu_*$ ($0<\nu_*<1$) in the Hartree-Fock approximation
\cite{Imo,Mae} 
and corresponds to the striped Hall 
state whose density is uniform in $y$ direction and periodic 
with a period $r_s$ in $x$ direction\cite{Mae}. 
The one-particle energy $\epsilon_{\rm HF}({\bf p})$ 
depends only on $p_y$ in this self-consistent solution. 
The Fermi sea and corresponding charge density distribution in $x$-$y$ 
space are sketched in Fig.~1. 
The period of stripe is $r_s$. 
The electric current flows along each stripes. 
Note that the true charge density and current density 
distribution in the mean field theory 
are not sharp as shown in the figure but fuzzy\cite{Mae}. 

\begin{figure}
\begin{center}
\includegraphics{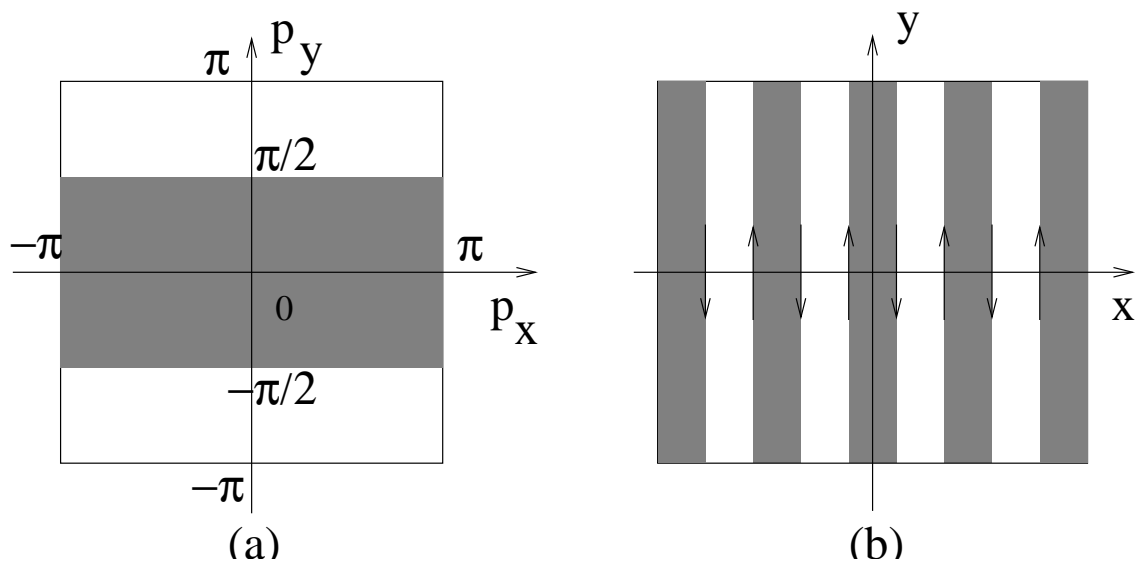}
\end{center}
\caption{(a) Fermi sea for the striped Hall state at the half-filling. 
(b) Charge density and electric current for the striped Hall state at 
the half-filling. }
\label{boring figure}
\end{figure}

Hartree-Fock energy per area for the striped Hall state is calculated as
\begin{equation}
E_{\rm HF}=\langle0\vert H^{(l)}\vert0\rangle/L^2
={1\over2}\int_{-\pi/2}^{\pi/2}{dp_y\over2\pi}\epsilon_{\rm HF}(p_y),
\end{equation}
where $L^2$ is an area of the system. 
$E_{\rm HF}$ depends on the period of the stripe $r_s$ and the 
optimal value of $r_s$ is determined by minimizing $E_{\rm HF}$. 
For $\nu=2+1/2$ and $3+1/2$, optimal values are $r_s=2.47$ and 
$2.88$, respectively. 

The striped Hall state obtained above  
breaks the magnetic translation symmetry in $x$ direction which 
corresponds to $Q_Y$. 
Using the commutation relation (\ref{com}), we can prove the 
Goldstone theorem for the striped Hall state\cite{SMA}. 
The theorem states that the NG mode appears at low energies and 
couples with the charge density operator. 
In the next section, we investigate the spectrum of the NG mode using 
the single mode approximation. 

\section{Single mode approximation}

We calculate the spectrum for a neutral collective 
excitation at the half-filled third and fourth Landau level 
using the single mode approximation. 
The single mode approximation is successful in the FQHS because 
the backflow problem is absent for the electron states projected 
to the Landau level\cite{Gir}. 
Projected density operator $\rho({\bf k})$ is written as 
$e^{-k^2/8\pi}L_l(k^2/4\pi)\rho_*({\bf k})$, where 
\begin{equation}
\rho_*({\bf k})=\int_{\rm BZ}{d^2p\over(2\pi)^2}b_l^\dagger({\bf p})
b_{l}({\bf p}-\hat{\bf k})e^{-{i\over4\pi}{\hat k}_x(2p_y-{\hat k}_y)}.
\end{equation}
Using $\rho_*$, the Hamiltonian $H^{(l)}$ is written as
\begin{equation}
H^{(l)}={1\over2}\int{d^2k\over(2\pi)^2}\rho_*({\bf k})
v_l(k)\rho_*(-{\bf k}),
\end{equation}
where $v_l(k)=e^{-k^2/4\pi}(L_l(k^2/4\pi))^2 2\pi q^2/k$ for 
the Coulomb potential. 
It is well-known that the density operators projected to the 
Landau level are non-commutative, that is,
\begin{equation}
[\rho_*({\bf k}),\rho_*({\bf k}')]=-2i\sin\left(
{{\bf k}\times{\bf k}'\over4\pi}\right)
\rho_*({\bf k}+{\bf k}').
\label{rho}
\end{equation}

The variational excited state is defined by $\vert{\bf k}\rangle=
\rho_*({\bf k})\vert{\rm stripe}\rangle$ and 
the variational excitation energy $\Delta({\bf k})$ is written as
\begin{eqnarray}
\Delta({\bf k})&=&{\langle{\bf k}\vert (H_l-E_0)\vert{\bf k}\rangle
\over\langle{\bf k}\vert{\bf k}\rangle}=
{f({\bf k})\over s({\bf k})},\nonumber\\
f({\bf k})&=&\langle0\vert [\rho_*(-{\bf k}),[H_l,\rho_*({\bf k})]]
\vert0\rangle/2N_e^*,\\
s({\bf k})&=&\langle0\vert\rho_*(-{\bf k})\rho_*({\bf k})
\vert0\rangle/N_e^*,\nonumber
\end{eqnarray}
where $E_0$ is a ground state energy, $N_e^*$ is a electron number 
in the l th Landau level, and 
$s({\bf k})$ is the so-called static structure factor. 
To derive these expressions, we use the relation $f(-{\bf k})=f({\bf k})$ 
and $s(-{\bf k})=s({\bf k})$ due to $\pi$ rotation symmetry. 
Using the commutation relation (\ref{rho}), $f({\bf k})$ is written as 
\begin{equation}
f({\bf k})=2\int{d^2k'\over(2\pi)^2}v_l(k')\sin^2
\left({{\bf k}'\times{\bf k}\over4\pi}\right)\{s({\bf k}+{\bf k}')
-s({\bf k}')\},
\end{equation}
where $v_l(k)=e^{-k^2/4\pi}(L_l(k^2/4\pi))^2 2\pi q^2/k$. 
Therefore the variational excitation energy is calculable if we know 
the static structure factor $s({\bf k})$. 
For the mean field state (\ref{mfs}) with $\nu_*=1/2$, 
$s({\bf k})$ becomes
\begin{eqnarray}
s({\bf k})&=&2\int_{\rm BZ}{d^2 p\over(2\pi)^2}
\theta(\mu-\epsilon_{\rm HF}(p_y))
\{1-\theta(\mu-\epsilon_{\rm HF}(p_y-{\hat k}_y))\}\\
&&+2 \sum_{N_x}{(2\pi)^2}\delta({\hat k}_x+2\pi N_x)
\delta({\hat k}_y)
\left({\sin(\pi N_x/2)\over \pi N_x}\right)^2.
\nonumber
\end{eqnarray}
The first term in $s({\bf k})$ behaves as $\vert k_y\vert/\pi r_s$ 
at small $k_y$ and periodic in $k_y$ direction with 
a period $2 \pi r_s$. 
The analytic form for $f({\bf k})$ are given by using 
Fourier series expansion as 
\begin{eqnarray}
f({\bf k})&=&2\sum_{n={\rm odd}}\{v_{\rm HF}(\sqrt{({2\pi n\over 
r_s}+k_x)^2+k_y^2)}-v_{\rm HF}({2\pi n\over r_s})\}
\left({\sin{nk_y\over 2r_s}\over\pi n}\right)^2,\\
v_{\rm HF}(k)&=&v_l(k)-
\int{d^2r}v_l(2\pi r)e^{i{\bf k}\times{\bf r}}.\nonumber
\end{eqnarray}
The numerical results for the energy spectrum $\Delta$ for $\nu=l+1/2$, 
$l=2$ and $3$ are shown in Figs.~2 and 3, respectively. 

\begin{figure}
\begin{center}
\includegraphics{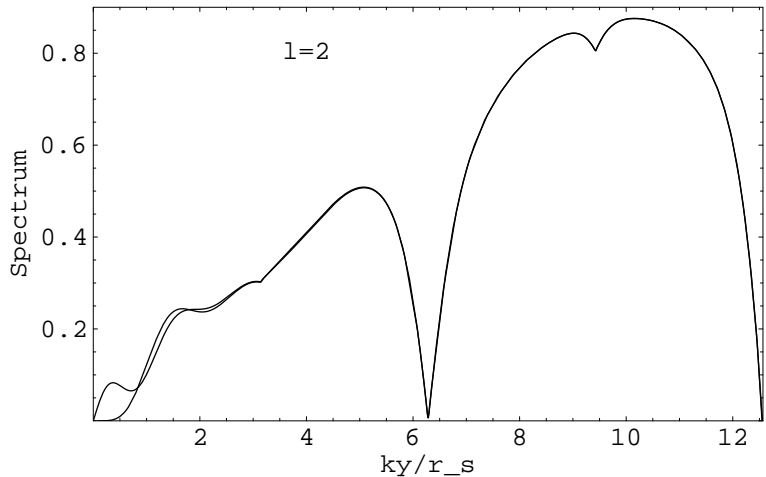}
\end{center}
\caption{Energy spectrum $\Delta$ at $0<r_s k_y<4\pi$ 
for $k_x=0$ and 1 (linear dispersion at $k_y=0$), 
$\nu=2+1/2$ in the single mode approximation. 
The unit of $\bf k$ is $a^{-1}$ and the unit of spectrum is 
$q^2/a$. The same unit is used in Fig.~3. }
\label{boring figure2}
\end{figure}

\begin{figure}
\begin{center}
\includegraphics{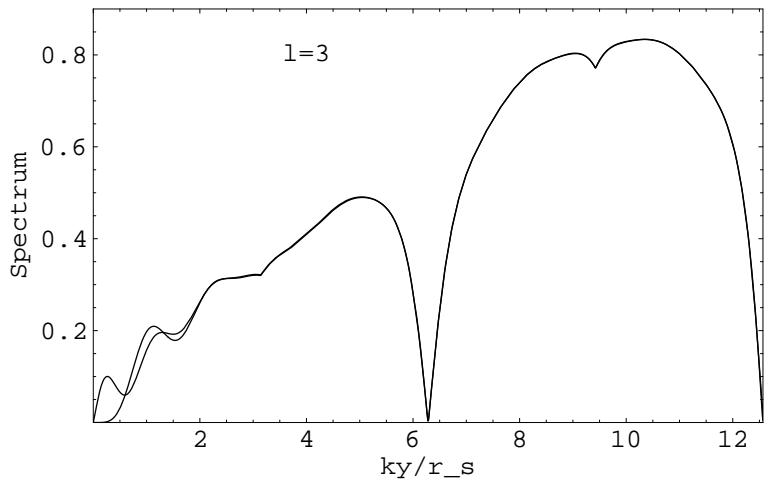}
\end{center}
\caption{Energy spectrum $\Delta$ at $0<r_s k_y<4\pi$ 
for $k_x=0$ and 1 (linear dispersion at $k_y=0$), 
$\nu=3+1/2$ in the single mode approximation. }
\label{boring figure3}
\end{figure}

As seen in these figures, the spectrum in $k_y$ direction 
resembles the liquid Helium spectrum with the phonon and roton minimum. 
The spectrum has a multiple line node at $k_y=2\pi r_s n$ 
($n$ is integer). 
The comparison to particle-hole excitation energy\cite{SMA} 
shows that the single mode approximation is good around 
$k_y=2\pi r_s n$. 
In a usual Fermi system, the single mode approximation seems not to work 
well because of the particle-hole excitation near the Fermi surface 
at low energies. 
In the present case, however, the low energy excitation near the 
Fermi surface is suppressed by the divergence of Fermi velocity 
at Fermi surface\cite{SMA}. 
Cusps also appear at $k_y=\pi r_s (2n+1)$. 

\section{summary}

We have studied neutral collective excitations in the striped Hall state. 
We have obtained the neutral excitation spectrum 
in the striped Hall state using the single mode approximation. 
The spectrum has a new rich structure. 
The neutral collective mode includes the NG mode due to 
spontaneous breaking of magnetic translation symmetry. 
The spectrum is highly anisotropic, that is, the dispersion in $k_x$ 
direction is similar to that of FQHS and the dispersion in $k_y$ is 
similar to that of Liquid Helium. 
We hope these excitations will be observed in experiments for the 
evidence of the striped Hall state.

\section*{Acknowledgements}
Two of the authors (T. A. and N. M.) thank J. L. Birman, 
A. Dorsey, R. M. Lewis, W. Pan, L. Tevlin, K. Yang, and J. 
Zhu for useful discussions. 
This work was partially supported by the special Grant-in-Aid for 
Promotion of Education and Science in Hokkaido University provided by the 
Ministry of Education, Science, Sport, and Culture, Japan, and by 
the Grant-in-Aid for 
Scientific Research on Priority area of Research (B) 
(Dynamics of Superstrings and Field Theories, Grant No. 13135201) from 
the Ministry of Education, Science, Sport, and Culture, Japan, and by 
Nukazawa Science Foundation.


\section*{References}
\begin{thebibliography}{9}

\bibitem{Lilly}
Lilly M P, Cooper K B, Eisenstein J P, Pfeiffer L N,
and West K W 1999 Phys. Rev. Lett. 82 394
\bibitem{Du}
Du R R, Tsui D C, Stormer H L, Pfeiffer L N, Baldwin K W, 
and West K W 1999 Solid State Commun. 109 389
\bibitem{Kou}
Koulakov A A, Fogler M M, and Shklovskii B I 1996 Phys. Rev. Lett.
76 499; Fogler M M, Koulakov A A, and Shklovskii B I 1996 
Phys. Rev. B 54 1853
\bibitem{Moe}
Moessner  R and Chalker J T 1996 Phys. Rev. B 54 5006
\bibitem{Imo}
Ishikawa K, Maeda N, and Ochiai T 1999 Phys. Rev. Lett. 82 4292
\bibitem{Mae}
Maeda N 2000 Phys. Rev. B61 4766
\bibitem{Ao}
Aoyama T, Ishikawa K, and Maeda N 2002 Europhys. Lett. 59 444
\bibitem{SMA}
Aoyama T, Ishikawa K, Ishizuka Y, and Maeda N cond-mat/0202405
\bibitem{Ima}
Imai N, Ishikawa K, Matsuyama T, and Tanaka I 1990 Phys. Rev. B 42 
10610; Ishikawa K, Maeda N, Ochiai T, and Suzuki H 1999 Physica E 
4 37; Ishikawa K and Maeda N cond-mat/0102347 
(unpublished); 2001 Physica B 98 159 
\bibitem{Gir}
Girvin S M, MacDonald A H, and Platzman P M 1986 Phys. Rev. B 
33 2481
\end{thebibliography}
\end{document}